\documentclass[aps,twocolumn,showpacs,superscriptaddress,amsmath,amssymb,nofootinbib]{revtex4-1}
\usepackage{graphicx}
\usepackage{chemist}
\usepackage[utf8]{inputenc}
\usepackage{hyperref}

\usepackage{multirow}
\usepackage{color}
\usepackage{ulem}
\usepackage{xspace}
\usepackage{gensymb}  
\usepackage{footmisc}

\DeclareGraphicsExtensions{.jpg, .png , .pdf, .bmp}

\definecolor{darkgreen}{rgb}{0.3,0.6,0.3}

\begin{document}

\title{Tuning bimetallic catalysts for a selective growth of SWCNTs }

\author{Salom\'e Forel$^{\ast}$}
\affiliation{Laboratoire de Physique des Interfaces et des Couches Minces, CNRS, Ecole Polytechnique, Universit\'e Paris Saclay, 91128, Palaiseau Cedex, France}

\author{Alice Castan$^{\ast}$}
\affiliation{Institut de Chimie Mol\'eculaire et des Mat\'eriaux d'Orsay, CNRS, Universit\'e Paris-Sud/Paris Saclay, 15 rue Georges Clemenceau, Orsay, France}
\affiliation{Laboratoire d'Etude des Microstructures, ONERA-CNRS, BP 72, 92322 Châtillon Cedex, France}

\author{Hakim Amara$^{\ast}$}
\affiliation{Laboratoire d'Etude des Microstructures, ONERA-CNRS, BP 72, 92322 Châtillon Cedex, France}

\author{Ileana Florea}
\affiliation{Laboratoire de Physique des Interfaces et des Couches Minces, CNRS, Ecole Polytechnique, Universit\'e Paris Saclay, 91128, Palaiseau Cedex, France}

\author{Fr\'ed\'eric Fossard}
\affiliation{Laboratoire d'Etude des Microstructures, ONERA-CNRS, BP 72, 92322 Châtillon Cedex, France} 

\author{Laure Catala}
\affiliation{Institut de Chimie Mol\'eculaire et des Mat\'eriaux d'Orsay, CNRS, Universit\'e Paris-Sud/Paris Saclay, 15 rue Georges Clemenceau, Orsay, France}

\author{Christophe Bichara}
\affiliation{Aix Marseille Universit\'e, CNRS, Centre Interdisciplinaire de Nanoscience de Marseille, Campus de Luminy, Case 913, F-13288, Marseille, France.}

\author{Talal Mallah}
\affiliation{Institut de Chimie Mol\'eculaire et des Mat\'eriaux d'Orsay, CNRS, Universit\'e Paris-Sud/Paris Saclay, 15 rue Georges Clemenceau, Orsay, France}

\author{Vincent Huc}
\affiliation{Institut de Chimie Mol\'eculaire et des Mat\'eriaux d'Orsay, CNRS, Universit\'e Paris-Sud/Paris Saclay, 15 rue Georges Clemenceau, Orsay, France}

\author{Annick Loiseau}
\affiliation{Laboratoire d'Etude des Microstructures, ONERA-CNRS, BP 72, 92322 Châtillon Cedex, France} 

\author{Costel-Sorin Cojocaru}
\affiliation{Laboratoire de Physique des Interfaces et des Couches Minces, CNRS, Ecole Polytechnique, Universit\'e Paris Saclay, 91128, Palaiseau Cedex, France}

\footnotetext{\textbf{corresponding author :} 
salome.forel@polytechnique.edu; \\alice.castan@onera.fr; hakim.amara@onera.fr}


\begin{abstract}

Recent advances in structural control during the synthesis of SWCNTs have in common the use of bimetallic nanoparticles as catalysts, despite the fact that their exact role is not fully understood. We therefore analyze the effect of the catalyst's chemical composition on the structure of the resulting SWCNTs by comparing three bimetallic catalysts (FeRu, CoRu and NiRu). A specific synthesis protocol is designed to impede the catalyst nanoparticle coalescence mechanisms and stabilize their diameter distributions throughout the growth. Owing to the ruthenium component which has a limited carbon solubility, tubes grow in tangential mode and their diameter is close to that of their seeding nanoparticle. By using as-synthesized SWCNTs as a channel material in field effect transistors, we show how the chemical composition of the catalysts and temperature can be used as parameters to tune the diameter distribution and semiconducting-to-metallic ratio of SWCNT samples. Finally, a phenomenological model, based on the dependence of the carbon solubility as a function of catalyst nanoparticle size and nature of the alloying elements,  is proposed to interpret the results.

\end{abstract}

\maketitle

\section{Introduction}

During the last decades, single-walled carbon nanotubes (SWCNTs) have been the focus of an intense research effort, highlighting their exceptional mechanical, electronic, optical and thermal properties~\cite{Dresselhaus2000}. In the field of nanotechnology, what appears today as the weak link for further application of carbon nanotubes is the difficulty to obtain large quantities and high purity of a unique structure of carbon nanotubes. Indeed, according to its structure, a nanotube can be metallic or semiconducting with a large range of gaps, depending on its diameter. Up to now only few chiralities have been obtained through selective synthesis ((6,5)~\cite{Bachilo2003, Miyauchi2004, Li2007, He2010}, (9,8)~\cite{Wang2013}, (6,6)~\cite{Sanchez2014}, (12,6)~\cite{Yang2014,Zhang2017a}, (16,0)~\cite{Yang2015}, (8,4)~\cite{Zhang2017a}, (14,4)~\cite{Yang2017}) but only the (6,6) synthesis has been reported to be 100 \% selective, albeit with a very poor yield. In the latter cases, the use of post-synthesis sorting is still needed. Importantly most of the cited syntheses focused on the use of bimetallic catalyst nanoparticles (NPs)~\cite{Bachilo2003, Miyauchi2004, Li2007, Wang2013, Yang2014,He2016}. Several hypotheses, supported by theoretical studies have been put forward to explain this chiral selectivity~\cite{Jourdain2013,Yang2016a,An_2016,Wang2018,He_2018,Penev_2018,HE2016243} but none of them have been unanimously accepted.\\

In the absence of a routine synthesis technique able to provide single-chirality SWCNT samples, various ways aiming at controlling the electronic properties of carbon nanotubes have been investigated~\cite{Harutyunyan2009b}, especially to favor the growth of semiconducting SWCNTs (s-SWCNTs) highly desired for applications in microelectronics. Two main routes are used: the selective etching of the metallic SWCNTs (m-SWCNTs) during the growth process~\cite{Hong2009,Yu2011, Qin2014, Ding2009, Li2013, Zhang2016}, or the burning of m-SWCNTs after integration into a device~\cite{Collins2001a}. Nevertheless, obtaining s-SWCNT-enriched samples is not sufficient for designing efficient devices: a fine-tuning of the diameter of the SWCNTs, which is inversely proportional to the gap, can greatly contribute to improving device performance~\cite{Zhang2016}. To this aim, the most common method relies on a fine control of the catalyst NP size where different assumptions are generally proposed. First, such control is constrained by coalescence mechanisms, which are highly activated at the temperature used in the chemical vapor deposition (CVD) process (600-1200\degree C). Indeed, many groups reported a global increase of nanotube diameters upon increase of the growth temperature, often explained in terms of catalyst coarsening leading to the disappearance of the smallest-sized catalyst particles, and subsequently of the small diameter SWCNTs~\cite{Wang2006, Tanioku2008, Li2009, Loebick2010}. This phenomenon, although likely to appear, was rarely demonstrated. The control of nanotube diameters is also explained by selective activation or deactivation of catalyst particles through their size. This selective activation is reportedly driven by the chemical interaction between carbon and the metal of the particle which strongly depends both on the catalyst size/nature and the growth temperature~\cite{Lu2006,Picher2009, Picher2011, Diarra2012a, Magnin2015}. 

Whatever the hypothesis put forward, these reported mechanisms imply that catalyst NP size and SWCNT diameter are very closely linked, which is not always the case. Indeed, two growth modes can be envisioned during a given growth event: a tangential growth mode where the diameter of the NP and the resulting nanotube are similar, and a perpendicular mode where they are no longer related~\cite{Fiawoo2012}. Moreover, for iron, nickel or cobalt  catalysts, these modes have been shown to depend on the wetting or dewetting properties of the carbon wall with respect to the catalyst surface, which are themselves driven by the fraction of carbon dissolved inside the catalyst~\cite{Amara2017, He2018}. \\

It is therefore obvious that the control of the chemical state of the catalyst NP in interaction with carbon precursors is a key factor for producing nanotubes with defined diameters. As a result, our approach relies first on favoring a tangential growth mode so that SWCNT diameters are governed by catalyst sizes, and suppressing catalyst coalescence mechanisms to get a good control of their sizes. In this context, we conducted a systematic study of our growth process by modulating the carbon solubility properties in the catalyst NPs by using different chemical compositions (FeRu, CoRu and NiRu) whose thermodynamic properties with respect to carbon are different, and performing growths at various temperatures. Under these conditions, a better understanding of the SWCNT growth mechanism is possible, resulting in more selective syntheses, with a better diameter control coupled with an increase of the semiconducting-to-metallic ratio.

\section{Results and discussion}

\subsection{Nanoparticle analysis}

For this study, we adapted the chemical composition of the catalyst in order to obtain catalysts with different chemical behaviors with respect to carbon, and to favor a tangential growth mode. We set our choice of catalysts of iron, cobalt and nickel mixed with ruthenium. Iron, cobalt and nickel metals have been chosen for their well known ability to seed the growth of SWCNTs and for their different chemical affinities for carbon. Indeed, as a general trend for transition metals, the carbon affinity decreases from the left to the right of the periodical table. In the bulk, carbon solubility is found to be around 0.3-0.5 at.~\% in Ni, 1.2 at.~\% in fcc-Co at 700\degree C, and around 4 at. \% for $\gamma$-Fe at 800\degree C~\cite{Jourdain2013}. Alloying these metals with ruthenium is expected to lower the carbon solubility limit (at 700\degree C carbon solubility in bulk ruthenium is around 0.15 at. \%~\cite{Yang2010}) and therefore to favor the tangential growth mode according to~\cite{He2018} and to increase the melting temperature~\cite{Hallstrom2004,Massalski1990} such as to maintain the alloy NPs in a solid state at the CVD growth temperatures. We therefore anticipate that employing these alloys foster a tangential growth mode on a solid catalyst. 

\begin{figure}[h]
\centering
	\includegraphics[width=1\linewidth]{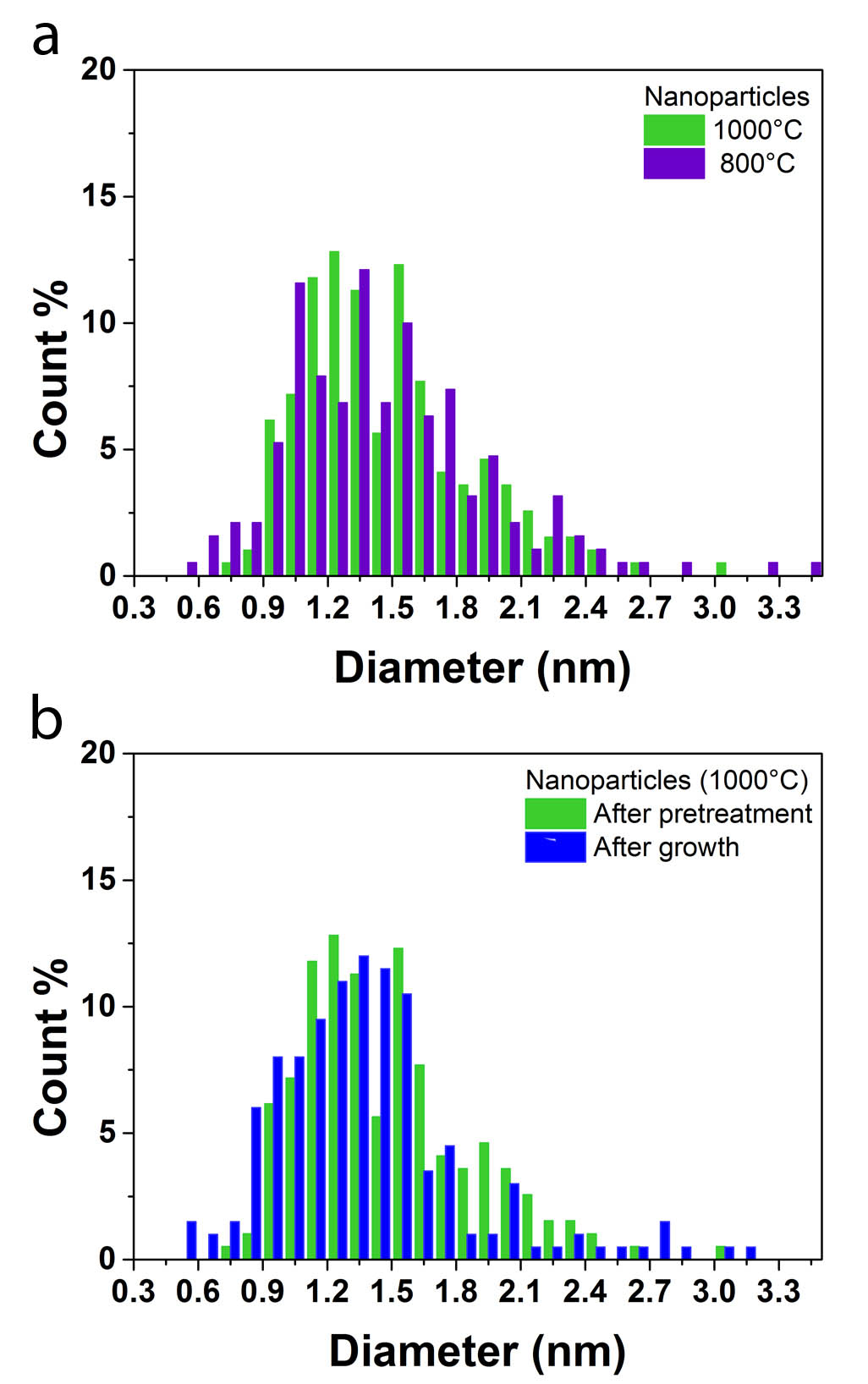}	
	\caption{(a) Size distribution histograms of FeRu NPs after pretreatment at 800\degree C (purple), and 1000\degree C (green),  (b) Size distribution histograms of FeRu NPs after the 5-minute pretreatment (green), and after growth (blue) at 1000\degree C (statistics on 200 NPs).}
\label{fig:figure1}
\end{figure}

 To produce bimetallic NPs (Fe-Ru, Co-Ru and Ni-Ru) with predefined compositions, we used a new SWCNT synthetic processes that we recently reported~\cite{Castan2017}, based on Prussian blue analogs (PBAs)~\cite{Catala2017} as catalyst precursors (see Methods section). This approach was shown to produce truly alloyed NPs with controlled and tunable chemical composition and also uniform size around 1.5 nm well-adapted to the growth of SWCNTs (see figure \ref{fig:figure1}, S1 and S2).

On the path to analyzing the effect of the growth temperature on the SWCNT growth, one of the major obstacles is related to coarsening and coalescence phenomena, such as the Oswald ripening~\cite{Voorhees1985} which is a thermally activated process. As previously described, many studies in the literature reported an increase of the as-grown SWCNT diameters at higher temperature due to an increase of the mean catalyst size induced by a coarsening mechanism~\cite{Wang2006, Tanioku2008, Li2009, Loebick2010}. In the proposed synthesis method, the activation of the coalescence mechanism is expected to be highly impeded. First, thanks to an appropriate surface chemical functionalization of the support substrate, the catalyst density has been shown to be limited to one monolayer.~\cite{Castan2017} Second, using highly activated hydrogen induces surface defects on the substrate with high-trapping energy over the substrate~\cite{Jeong2008}. These defects limit the surface diffusion of the catalyst metal atoms and increase the pinning effect of the forming NPs~\cite{Castan2017, Jeong2008, Bouanis2014}.

In order to verify such unusual limited coalescence, we compared, using statistical transmission electron microscopy (TEM) measurements (around 200 NPs counted for each sample), the size distribution of the catalyst NPs after the pretreatment step preceding their exposure to the carbon gas source at 800 and 1000\degree C. As an example, the case of FeRu is discussed here whereas results for other alloys can be found in figure S1 of the supporting information. Figure~\ref{fig:figure1}a presents the size distribution histograms of FeRu NPs after the five-minute pretreatment conducted in the CVD reactor at 800\degree C and 1000\degree C, respectively. The obtained size distributions are very similar. In both cases, small NPs are mostly present over the surface, 57 \% of counted NPs are below 1.5 nm at 800\degree C and 56 \% at 1000\degree C. These measurements let us assume that during the SWCNT synthesis process, after the 5-minute pretreatment, when the carbon feed is introduced, the same population of catalytic NPs is available for subsequent SWCNT growth. Furthermore to ensure that the coalescence does not occur after the introduction of methane and during the 30 minutes of SWCNT growth, we also compared the catalyst size distributions after the pretreatment step and respectively after the SWCNT growth. As presented in figure ~\ref{fig:figure1}b, for the synthesis process at 1000\degree C, the highest temperature in this study, and for which it is reasonable to expect the highest activation of the coalescence mechanism, the catalyst diameter distributions are very similar after the pretreatment step and after the half-hour growth and only extremely limited traces of coalescence can be observed. Similar results for all three catalysts but for a synthetic process at 800\degree C have been already demonstrated~\cite{Castan2017}.

\subsection{Tube/catalyst interaction}

\begin{figure}
\centering
	\includegraphics[width=\linewidth]{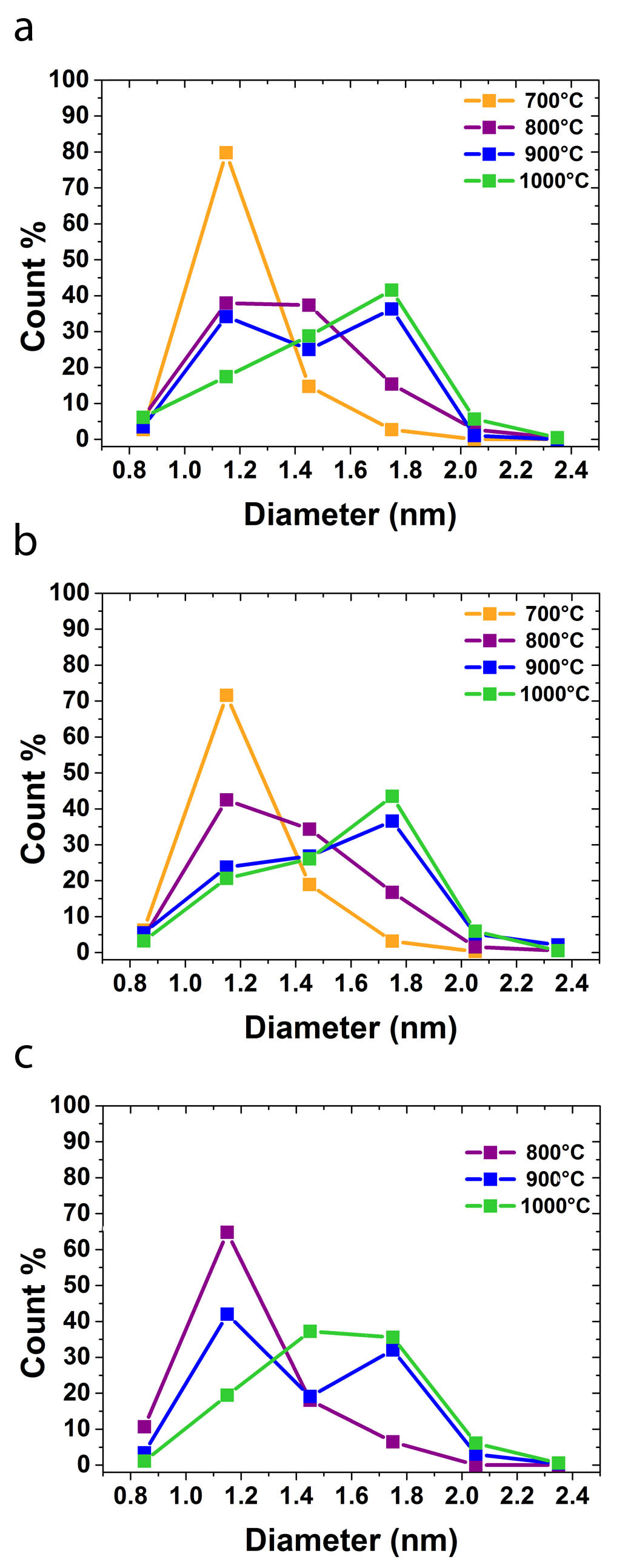}	
	\caption{Diameter distributions obtained through Raman spectroscopy analysis for growths from (a) FeRu catalysts, (b) CoRu catalysts, and (c) NiRu catalysts. The point is placed in the middle of each diameter bin (0.3 nm wide).}
\label{fig:figure2}
\end{figure}

We now focus on the diameter distribution of the tubes grown by performing systematic investigations of the growth temperature influence for the three types of nanoalloy catalysts at 700\degree C, 800\degree C, 900\degree C and 1000\degree C, respectively. The SWCNT diameter distributions were investigated through Raman spectroscopy with four excitation wavelengths (see Methods section). Figure ~\ref{fig:figure2} represents the results obtained for one batch of samples (see figure S3 in SI). The reproducibility of the results was checked and is available in SI (figure S4) for the FeRu system. 

Contrary to the particle diameters, the detected SWCNT diameters are clearly temperature-dependent and following the same trend, whatever the catalyst (see figures ~\ref{fig:figure2}a-c), even if one can note that nanotubes could not be grown at 700\degree C for the NiRu catalyst system, and were only observed at 800\degree C and above, this point will be discussed later. At low temperature (700\degree C for the FeRu and CoRu systems, or 800\degree C for the NiRu system), a majority of small diameter SWCNTs are synthesized and very few larger diameter nanotubes were observed.
For instance, for each of the three catalysts, when the growth is performed at the lowest temperature (700\degree C for FeRu and CoRu catalyst or 800\degree C for NiRu catalyst), more than 65 \% of the counted SWCNTs have diameters below 1.3 nm, and fewer than 6 \% have diameters above 1.6 nm. Upon temperature increase, the number of small-diameter SWCNTs decreases whilst the number of larger diameter SWCNTs increases. At 1000\degree C, less than 23 \% of the nanotubes are found to have diameters below 1.3 nm and more than 40 \% above 1.6 nm.  

\begin{figure}[h]
\centering
	\includegraphics[width=\linewidth]{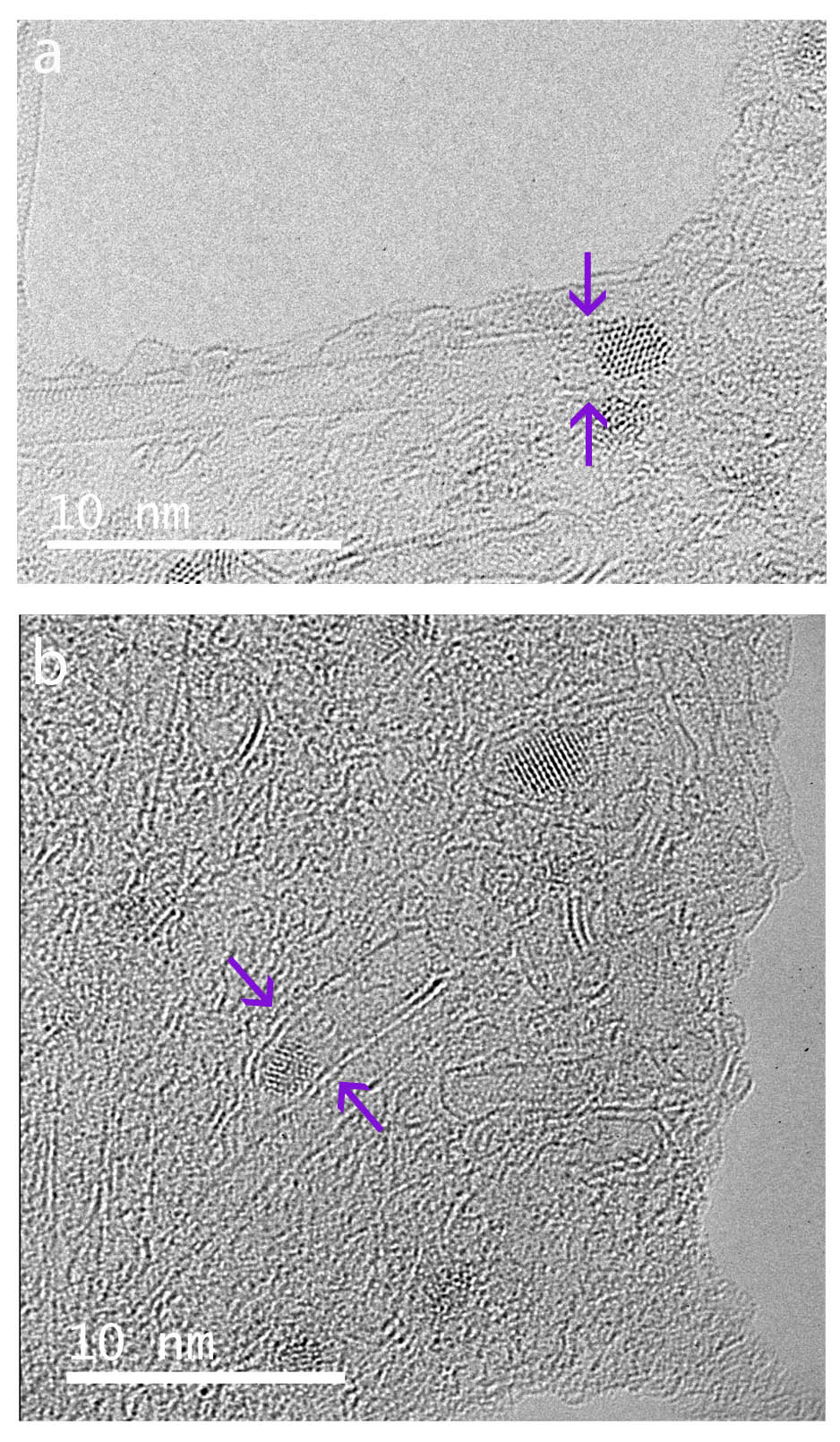}	
\caption{High resolution TEM (HRTEM) images of nanotubes attached to their corresponding catalyst NPs in the cases of growth from (a) the CoRu catalyst at 800\degree C  and (b) from the FeRu catalyst at 900\degree C (bottom), both in a tangential growth configuration.}
\label{fig:figure3}
\end{figure}

The next step was to examine the growth mode in order to determine the origin of the temperature dependence of SWCNT diameters. In all cases where a nanotube was seen attached to its catalyst NP during our TEM investigations, the observed growth mode was systematically the tangential  type independently of the NP diameter (see figure ~\ref{fig:figure3}). This finding reasonably validates our approach in designing a catalyst to privilege a tangential growth by tuning the carbon concentration in the catalyst. Importantly, if we consider the diameters of the NP and the attached nanotube to be close to equal, the small NPs appear to be less active for growth at 1000\degree C, while bigger NPs are less active at low temperature (see figure S5 in SI). To confirm a selective activation of the available NPs with temperature, we have performed the SWCNT growth  at 800\degree C, but starting with catalyst NPs pretreated five minutes at 1000\degree C (see figure S6 in SI). The SWCNT diameter distribution is similar to those obtained for a full synthesis procedure at 800\degree C, supporting a selective activation well related to the growth temperature.

We can therefore conclude that catalyst NPs available for synthesis are roughly in the same size range whatever the growth temperature, but that they are selectively activated according to their size, leading to SWCNTs of different diameter ranges as a function of temperature.

\subsection{Phenomenological model}

\begin{figure*}[ht]
	\includegraphics[width=1\linewidth]{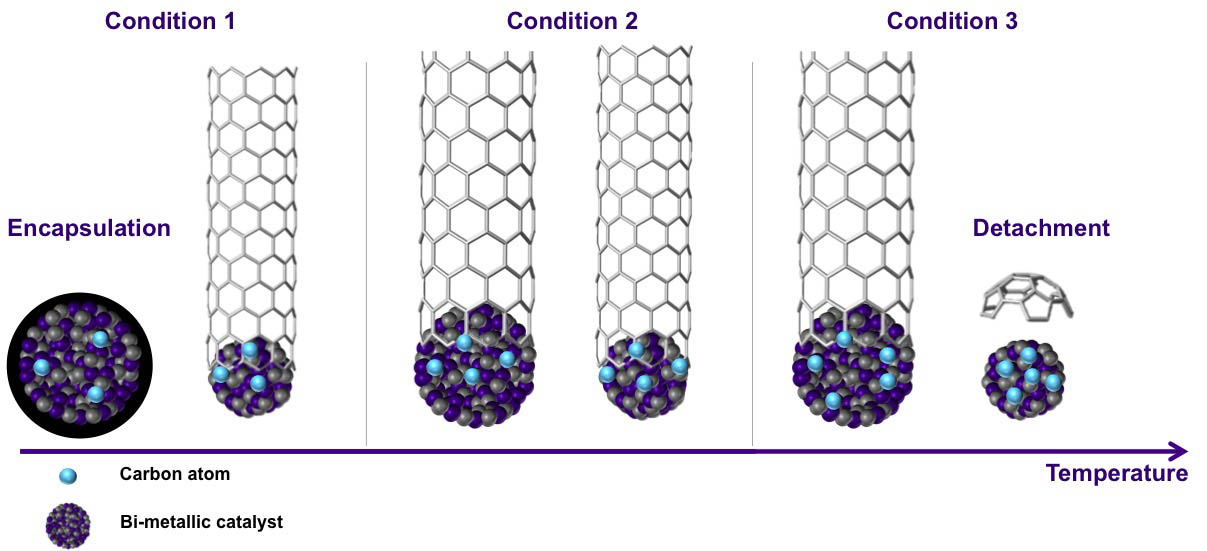}	
	\caption{Schematic view of the proposed mechanism of selective activation of catalyst NPs based on growth temperature and NP size.}
\label{fig:figure4}
\end{figure*}
 
It is commonly admitted that SWCNT growth can start only after the carbon saturation of the corresponding catalyst NP has been reached~\cite{Shibuta2003, Zhao2005, He2015, Amara2017}. In the particular context of nanotube growth, theoretical studies forecast an increase of carbon solubility with the decrease of the NP size which can affect the growth mechanism~\cite{Ding2006, Diarra2012a, Diarra2012b, Magnin2015}. Experimentally, Picher and co-workers~\cite{Picher2009, Picher2011} reported that at high growth temperatures in the low pressure domain,  small catalyst NPs were not activated due to a higher supersaturation limit as compared to that of the larger particles in agreement with~\cite{Navas2017}. This hypothesis could be used to explain the lack of small diameter SWCNTs observed at high temperature, but does not seem to be very realistic within the frame of our specific growth conditions. Indeed, the partial pressure of methane is relatively high (around 27 mbar) and the synthesis time is relatively long (30 minutes), it seems difficult to envision a possible lack of sufficient carbon feed of the catalyst as a limiting factor. Besides, recently published calculations on SWCNT growth from NPs with various carbon concentrations highlighted its influence  on the nanotube's nucleation~\cite{Aguiar2017}. In the case of a catalyst with a low carbon concentration, a strong adhesion between the initially formed carbon $sp^{2}$ cap and the NP is expected leading to their poisoning. On the contrary, it is found that the high-carbon concentrations metal NPs present an insufficient adhesion towards the nanotube, inducing the SWCNT growth termination.\\

On the basis of all these considerations, we propose a growth process sketched in a simplified scheme in figure ~\ref{fig:figure4} to account for our experiments. The key point is to distinguish two kinds of NPs depending on their size: "small" versus "large" NPs. Using this crude separation, we propose to explain the catalyst size selectivity as follows. At low temperatures, the carbon fraction of larger particles is low. According to calculations done in~\cite{Aguiar2017}, this lack favors a strong adhesion of the $sp^{2}$ layer on the catalyst surface which leads to the encapsulation and deactivation of the NP (condition 1 in figure ~\ref{fig:figure4}). On the contrary due their higher carbon fraction, small catalyst NPs tend to favor the dewetting of the $sp^{2}$ cap formed on their surface and are capable of initiating the SWCNT growth. When the growth temperature increases, carbon solubility in larger NPs increases and  gets high enough to allow the lift-off of the  $sp^{2}$ cap (condition 3 in figure ~\ref{fig:figure4}). In turn, the carbon solubility in smaller NPs also increases but too much, and can induce a detachment of the carbon cap so rapid that it prevents the nanotube growth initiation (condition 3 in figure ~\ref{fig:figure4}). In the intermediate temperature range, the carbon solubility limit of both smaller and bigger catalyst NPs are propitious to nanotube growth initiation (condition 2 in figure ~\ref{fig:figure4}). \\

 The present phenomenological model of the temperature dependence of carbon solubility in NPs relatively to their size  accounts well for the general trend regarding the evolution of SWCNT diameter distributions observed in figure ~\ref{fig:figure2}. Indeed, in view of the data for the FeRu catalyst system, we first observed at low temperature the growth of small SWCNTs (in line with condition 1 in figure ~\ref{fig:figure4}). Then, when the temperature increases, a combination of small and larger tubes is obtained (in line with condition 2 in figure ~\ref{fig:figure4}). And finally for the highest temperature a majority of large nanotube grown (in line with condition 3 in figure ~\ref{fig:figure4}). Few differences are observed between the growth from the FeRu and CoRu catalysts, but a clear difference appears with the NiRu  catalyst system where no SWCNT growth is observed at the lowest temperature 700\degree C. Referring back to our model, a lower carbon solubility in the NiRu catalyst as compared to the two other alloys can explain such a result. At the lowest tested temperature (700\degree C), carbon solubility in the NiRu NPs in the available diameter range is to low to induce a growth of SWCNTs. As the temperature increases, the carbon solubility increases and the NPs are gradually activated  through their size : first the smaller then the bigger ones, until the occurrence of the deactivation of the smaller size ones at high temperature (1000\degree C). However, assuming a temperature upshift of 100\degree C, between NiRu and (Fe/Co)Ru, SWCNT diameter distributions get quite comparable. This lower carbon solubility in NiRu is nicely consistent with the relative carbon solubility in the corresponding bulk materials, that is ten times lower in nickel as compared to cobalt and iron, the presence of ruthenium seems therefore significantly enhances this effect.
 
\subsection{Impact of the nanotube diameter distribution control on the SC/M ratio}

\begin{figure}[ht]
\centering
	\includegraphics[width=1\linewidth]{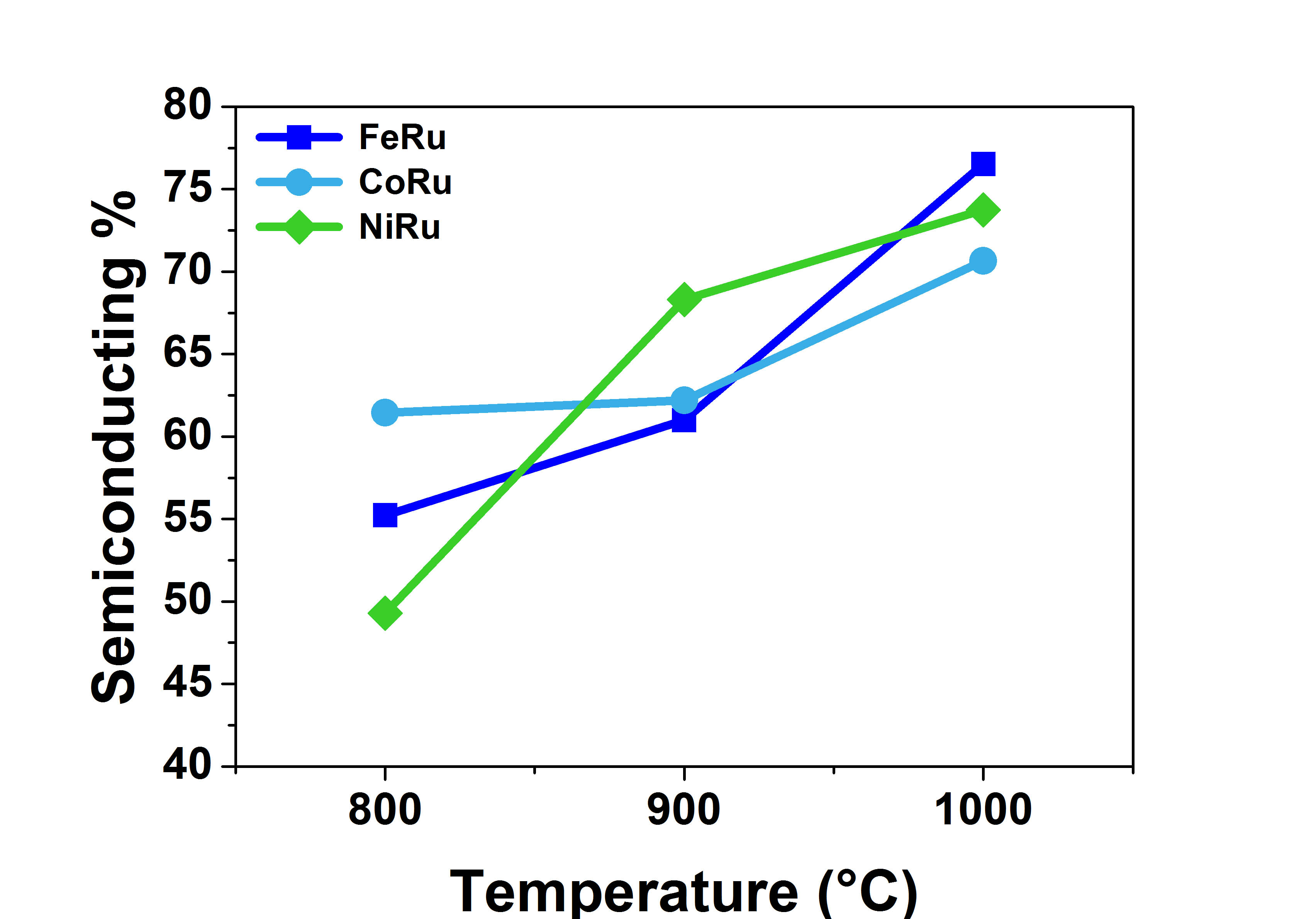}	
	\caption{Evolution of the percentage of s-SWCNTs determined by Raman spectroscopy analysis for the three catalysts at various growth temperatures.}
\label{fig:figure5}
\end{figure}

The ability to control and choose the diameter of the growing SWCNT previously discussed can now be exploited to favor a tunable enrichment of the semiconducting or metallic SWCNT populations in the samples. Indeed, it has been shown that the use of an etching gas to obtain s-SWCNT enrichment during CVD growth seems to be efficient only in a specific diameter range (typically above 1.4 nm)~\cite{Yu2011, Qin2014, Ding2009, Li2013, Zhang2016}. For smaller diameters ($<1.4$ nm), no selective etching of the m- and s-SWCNTs was observed. Li and co-workers reported a similar diameter dependence trend when using hydrogen as an etching gas~\cite{Li2013}. Here we show that our catalyst family is not only capable of allowing SWCNT growth with selective diameter distribution but also with selective semiconducting/metallic character. \\

To that end, we used Raman spectroscopy data to determine the SC/M ratio of our SWCNT samples (see methods section for more detail). As seen in figure ~\ref{fig:figure5}, we can clearly observe an increase in the percentage of s-SWCNTs with the temperature for all catalysts. This selectivity is explained first by the activated hydrogen used during the synthesis, which acts as an efficient etching agent, and second by the fact that at high temperature tube diameters range above 1.4 nm, the lower limit above which etching is efficient.

To prove the efficiency of our method for growing SWCNTs with a selectivity in diameter and SC/M ratio, we integrated the as-synthesized SWCNTs as a channel material in field effect transistors (FETs) (see figure S7 in SI for a schematic view of the device) and subsequently statistically analyzed their electrical characteristics. This measurement also allows cross-checking of the Raman spectroscopy measurements. As we performed the SWCNT growth experiments on \chemform{Si/SiO_2} wafers, the as-obtained SWCNTs can directly be integrated in bottom gate SWCNT-FET device structures without transfer steps (see Methods section). As very few SWCNTs were observed for the growth from the NiRu catalyst at 700\degree C, we chose to use nanotubes grown from the three catalysts types at two extreme growth temperatures, respectively 800\degree C and 1000\degree C. Over fifty devices were characterized for each starting catalyst type and growth temperature. We will focus the analysis on the On-current value and the Ion/Ioff ratio value for the as-fabricated device, before and after a breakdown process (see Methods section), which is intended to disconnect the metallic SWCNTs present in the FET channel.

\begin{figure*}
	\begin{center}
	\includegraphics[width=1\linewidth]{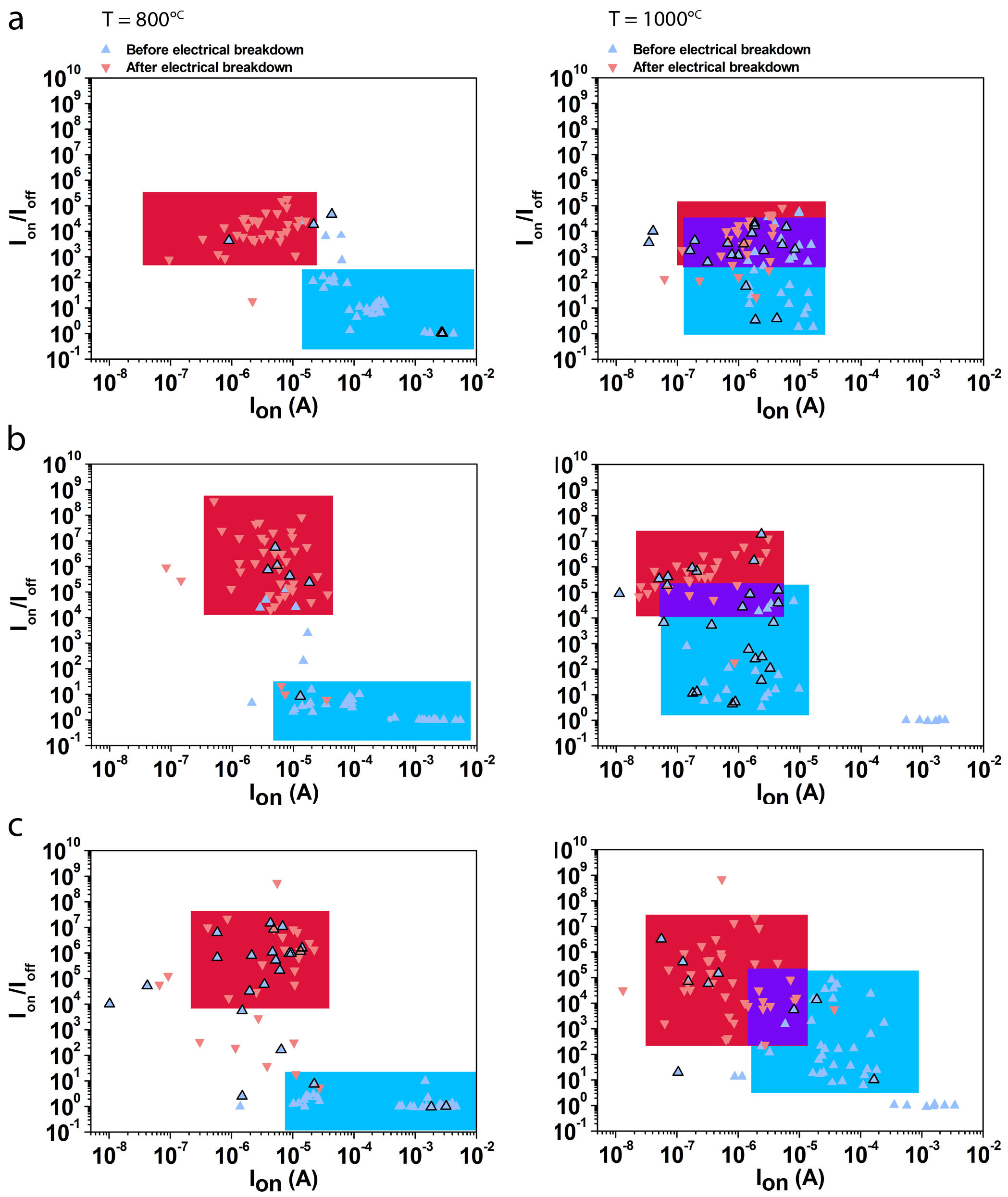}	
	\end{center}
	\caption{Ion/Ioff ratios of FET-devices fabricated with SWCNTs from (a) FeRu catalysts (b) CoRu catalysts and (c) NiRu catalysts, at (left) 800\degree C  and (right) 1000\degree C. The blue points correspond to the results obtained before any electrical breakdown while the red points correspond to the results obtained after the electrical breakdown. Black points indicate the transistors where the electrical breakdown process was found to be ineffective.}
\label{fig:figure6}
\end{figure*}

Before applying the breakdown process (blue points in figure ~\ref{fig:figure6}), for all devices, we observe a similar trend as a function of the SWCNT growth temperature, independent of the catalyst type. When SWCNTs grown at the lowest temperature are integrated in a transistor channel (figure ~\ref{fig:figure6} and figure S8), a majority of the devices are found to be ineffective (i.e. Ion/Ioff ratio lower than 10). For SWCNTs grown at high temperature (figure ~\ref{fig:figure6} and figure S8), however, few of the fabricated transistors are ineffective.

One can observe that electrical breakdown of all the measured transistors has only a limited effect on transistors integrating nanotubes grown at 1000\degree C. As shown in figure ~\ref{fig:figure6}, a very weak improvement of the Ion/Ioff ratio is obtained and the On-current stays relatively constant. We calculated the percentage of transistors presenting an improvement of Ion/Ioff ratio larger than 3 decades after the electrical breakdown (see figure S8 in SI). For instance, such improvement concerns 6 \% of transistors made with nanotube synthetized from FeRu at 1000\degree C which is not significant. We therefore conclude that only a weak proportion of m-SWCNTs connects source and drain, again confirming the s-SWCNT enrichment detected through Raman spectroscopy. On the contrary, the electrical breakdown appears to be very effective for transistors made from SWCNTs grown at 800\degree C. After the breakdown, for each of the transistors with 800\degree C grown nanotubes, the Ion/Ioff ratio increased and the On-current decreased. Here, an improvement of the Ion/Ioff ratio superior to 3 decades is around 28 \% for the transistors with SWCNTs grown at 800\degree C from FeRu (see figure S8 in SI). Such results are expected when the contribution of m-SWCNTs is suppressed. This result highlights a higher contribution of m-SWCNT to this type of transistors, consistent with the Raman characterization.

Further insights are expected to be retrieved after applying the breakdown process for disconnecting metallic conducting paths from the FET channels.  Particularly the electronic band gap of s-SWCNTs being diameter-dependent, the Off-current value is expected to increase as the nanotube's bandgap decreases (as the Schottky barrier height at contacts is expected to decrease and consequently unwanted tunneling carrier injection increases). Then, after electrical breakdown, transfer characteristics obtained for FET devices based on SWCNTs grown at 800\degree C should present higher Ion/Ioff ratios. This behaviour is indeed well visible in figure S9 which shows the Off-current obtained in all the measured transistors after electrical breakdown as a function of the Ion/Ioff ratio, confirming that smaller  SWCNTs have been obtained for this lower temperature growth. 
Finally, it is important to note that despite the simplicity of the transistor design, our synthesis method allows to easily obtain transistors with good Ion/Ioff ratios. Indeed, Ion/Ioff ratios larger than eight decades for the best ones, and around five decades on average have been obtained.  These performances are on the same order of magnitude as the best devices obtained using random percolating networks of SWCNTs as channel reported in the literature \cite{Brady2016,Laiho2017,Sun2011}. 

\section{Experimental section}

\subsection{Synthesis of the Prussian blue analogs (pre-catalyst)}

An aqueous solution containing the hexa-aquo complex \chemform{[M(H2O)]^6_{2+}} (5 mM) and cesium chloride (5 mM) (CsCl) (CoRu and NiRu system) or potassium chloride (KCl for FeRu system) (5mM), was added to an aqueous solution containing the potassium hexacyanometallate \chemform{K_4Ru(CN)_6} salt (5 mM). 

\subsection{SWCNT growth}

The synthetis method used in this work has been detailed in previous work~\cite{Castan2017}. Briefly, the surface of a  \chemform{SiO_2/Si} wafer was first covered with a self-assembled monolayer (SAM) of a pyridine-functionalized silane. Then, the bimetallic (FeRu, CoRu and NiRu) PBA NPs were assembled by coordination bonds on the pre-formed organic SAM. A 5-minute pretreatment under activated hydrogen in the CVD chamber allowed the reduction of the PBA NPs into bimetallic NPs, which is the effective catalyst for the SWCNT growth. The resulting alloy NPs were then used to catalyze SWCNT growth via \chemform{CH_4/H_2} double hot-filament chemical vapor deposition (d-HFCVD) (growth time is 30 minutes). For the present study, syntheses (pretreatment, and subsequent growth) were performed at four different temperatures (700\degree C, 800\degree C, 900\degree C, and 1000\degree C) for each catalyst. For each synthesis temperature, the synthesis of SWCNTs was performed on three samples (one of each catalyst) at the same time. Under this condition, we ensure to analyze the effect of the synthesis temperature on the SWCNT structure for a given catalyst or the effect of the catalyst chemical composition for a fixed growth temperature. 

\subsection{Electrode deposition}

UV-lithography was used to deposit palladium electrodes. First, the \chemform{SiO_2/Si} substrate with SWCNTs was cleaned in a sonication bath for three minutes in acetone, and isopropanol, subsequently. Then, photosensitive resin \chemform{SU_8} was spin-coated (30 seconds at 4000 rpm) and the wafer was annealed for 5 minutes at 110\degree C. A patterned quartz mask was used to allow selective UV-illumination. The sample was illuminated for 7 seconds at 10 mW by a UV-light. The sample was then immersed for 25 seconds in a solution of MF 319, where the UV-exposed resin was removed. Then, the sample was transferred to an evaporator system to undergo the deposition of 40 nm of palladium. Finally, the sample was placed in a boiling acetone solution for a few minutes, the remaining photoresin covered by metal, was removed. The wafer was then rinsed in a boiling isopropanol solution. 

\subsection{Electrical breakdown of m-SWCNTs}

We used the electrical breakdown method developed by IBM~\cite{Collins2001a} to selectively disconnect metallic SWCNTs. During the process the s-SWCNTs are protected by depleting them from their carriers, through the application of an adapted gate voltage to set the field effect transistor (FET) device in its Off state and then a high voltage is applied between source and drain. Here we fixed the gate voltage at 25 V and the voltage between source and drain varied gradually between 0 and 25 V. 

\subsection{Nanoparticle characterization}
 
Diameter distributions of the catalyst NPs were obtained through TEM imaging. Catalyst NPs after pretreatment or after growth are transferred onto a TEM grid according to a process previously described~\cite{Castan2017}. Then, TEM imaging was performed using an image-corrected FEI TITAN environmental TEM (ETEM) operating at 300 kV.

\section{Conclusions}

In this work, three bimetallic catalyst systems were studied, all of which were prepared by the same process and obtained with identical morphologies and distributions. The stability of the catalyst diameter distribution throughout the growth process was demonstrated by the TEM statistical analysis. A statistical analysis of the nanotubes' diameter distribution performed by multiwavelength Raman spectroscopy shows a global shift of distributions towards larger diameters with increasing temperature. Assuming a tangential growth mode, based on TEM images of SWCNT attached to catalyst nanoparticles after growth, we were  able to establish a link between the evolution of diameter distributions, not with an increase in the size of the catalyst NPs, but rather with a phenomenon of selective activation of different catalyst size populations in the same initial pool of available catalysts by tuning the growth temperature. We assume that this effect can be correlated with the difference in carbon solubility in NPs according to their size and chemical composition and propose a phenomenological model. Indeed, the role of ruthenium is twofold because it makes it possible to limit the carbon concentration in NPs and thus to promote a tangential mode. In addition, the carbon concentration in the NPs can be modulated to obtain a more precise control of the tubes produced. Further, using a Raman characterization of our samples, confirmed by a statistical analysis of FET devices, we have evidenced a temperature evolution of the SC/M ratio, that we analyse as being a direct consequence of the diameter tuning in our etching synthesis conditions.
Finally, with an appropriate choice of the growth temperature, the devices manufactured have shown performance at the state of the art. Our work is therefore a first step to better understand why bimetallic catalysts are so interesting for the growth of SWCNTs and how to adjust their properties in order to obtain a more selective synthesis. It provides a rational understanding of the key factors that determine SWCNT diameters, providing new elements to grow SWCNTs with defined chiralities.

\begin{acknowledgments}

The research leading to these results has received funding from the French Research Funding Agency under grant No. ANR-13-BS10-0015 (SYNAPSE project), the European Union Seventh Framework Programme (FP7/ 2007e2013) under grant agreement 604472 (IRENA project),  Tempos-NanoTEM and TEMPOS NanoMax under the reference ANR-10-EQPX-50 and the Chaire de Recherche Andr\'e Citroen (PSA AC3M) at Ecole Polytechnique (France). The authors thank the METSA research foundation for giving access to the Cs-corrected TEM of MPQ-Paris Diderot laboratory.

\end{acknowledgments}

\bibliographystyle{apsrev4-1}
\bibliography{biblio-swcnt-growth}

\end{document}


\begin{center}
\textbf{Electronic Supplementary Information}
\end{center}

\begin{center}
\LARGE{\textbf{Tuning bimetallic catalysts for a selective growth of SWCNTs}}
\end{center}

\noindent\large{Salom\'e Forel,\textit{$^{a}$} Alice Castan,\textit{$^{b,c}$} Hakim Amara,\textit{$^{b}$} Ileana Florea,\textit{$^{a}$} Fr\'ed\'eric Fossard,\textit{$^{b}$} Laure Catala,\textit{$^{c}$} Christophe Bichara,\textit{$^{d}$} Talal Mallah,\textit{$^{c}$} Vincent Huc,\textit{$^{c}$} Annick Loiseau,\textit{$^{b}$} and Costel-Sorin Cojocaru\textit{$^{a}$}} 

\footnotetext{\textit{$ ^{a}$~Laboratoire de Physique des Interfaces et des Couches Minces, CNRS, Ecole Polytechnique, 91128, Palaiseau Cedex, France ; E-mail: salome.forel@polytechnique.edu}}
\footnotetext{\textit{$ ^{b}$~Laboratoire d'Etude des Microstructures, ONERA-CNRS, UMR104, Universit\'e Paris-Saclay, BP 72, 92322 Chatillon Cedex, France ; E-mail: alice.castan@onera.fr, hakim.amara@onera.fr}}
\footnotetext{\textit{$ ^{c}$~Institut de Chimie Mol\'eculaire et des Mat\'eriaux d'Orsay, CNRS, Universit\'e Paris-Sud/Paris Saclay, 15 rue Georges Clemenceau, Orsay, France}}
\footnotetext{\textit{$ ^{d}$~Aix Marseille Universit\'e, CNRS, Centre Interdisciplinaire de Nanoscience de Marseille, Campus de Luminy, Case 913, F-13288, Marseille, France}}

\section{Size distribution}

\begin{figure}[ht]
	\includegraphics[width=\linewidth]{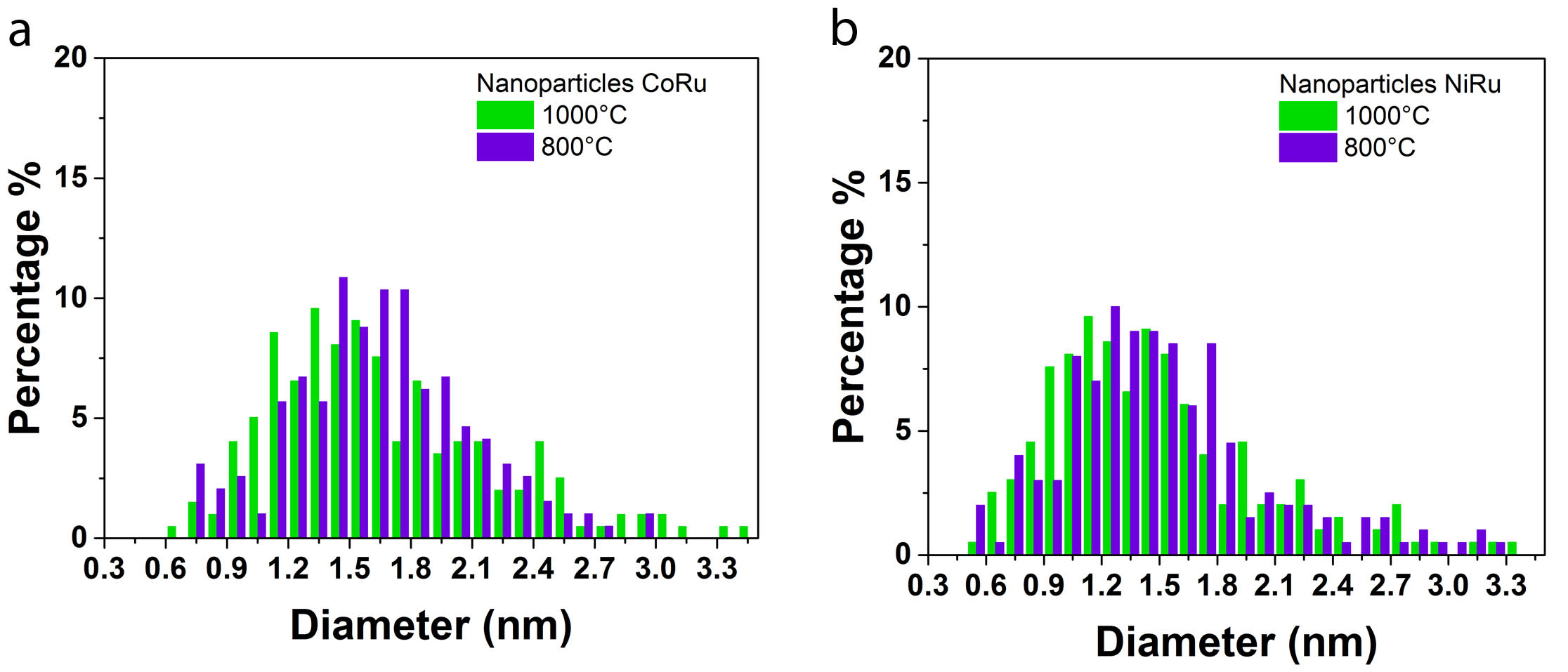}	
	\caption{Size distribution of NiRu and CoRu nanoparticles after pretreatement at 800 amd 1000\degree C.}
\end{figure}

\newpage

\newpage

\section{STEM-EDX study}

\begin{figure}[h!]
	\includegraphics[width=\linewidth]{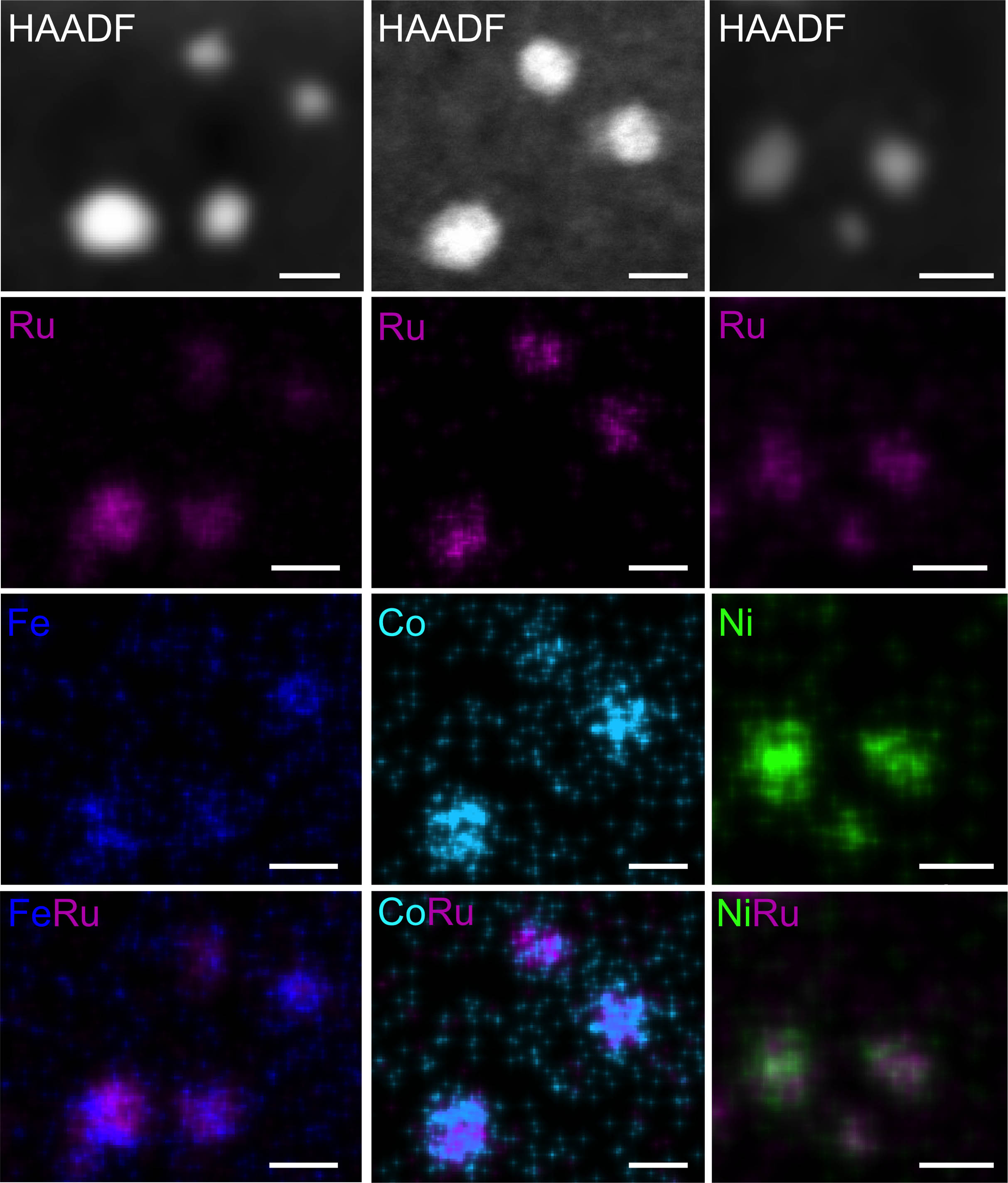}	
	\caption{HAADF-STEM images, and EDX-STEM chemical maps for the three systems (FeRu, CoRu, NiRu). For each image the scale bar is 3nm.}
\end{figure}

\newpage

\section{Methodology}

\begin{figure}[ht]
\begin{center}
	\includegraphics[width=\linewidth]{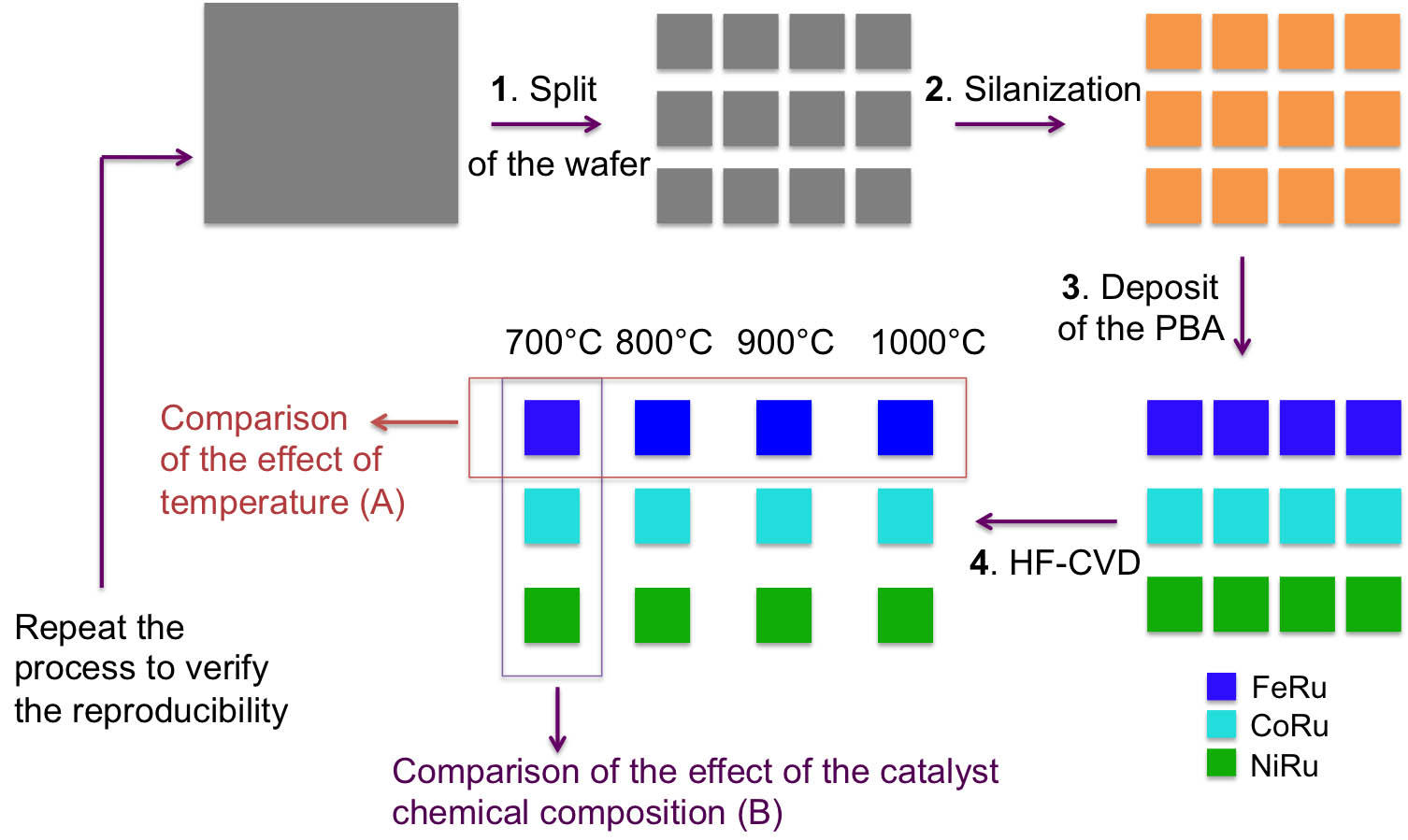}	
	\caption{Schematic view of the synthesis process in order to realized a parametric study.}
	\end{center}
\end{figure}

The SWCNT growth process (including pre-treatment and growth) is performed in the HF-CVD at 700\degree C 800\degree C, 900\degree C and 1000\degree C. For each synthesis temperature, the synthesis of SWCNTs are performed on three samples (one of each catalyst) at the same times. Under this condition, we can analyze the effect of the synthesis temperature on the SWCNT structure for a given catalyst (step A) or the effect of the catalyst chemical composition for a fixed growth temperature (step B).

\newpage

\section{Reproducibility}

\begin{figure}[ht]
	\includegraphics[width=\linewidth]{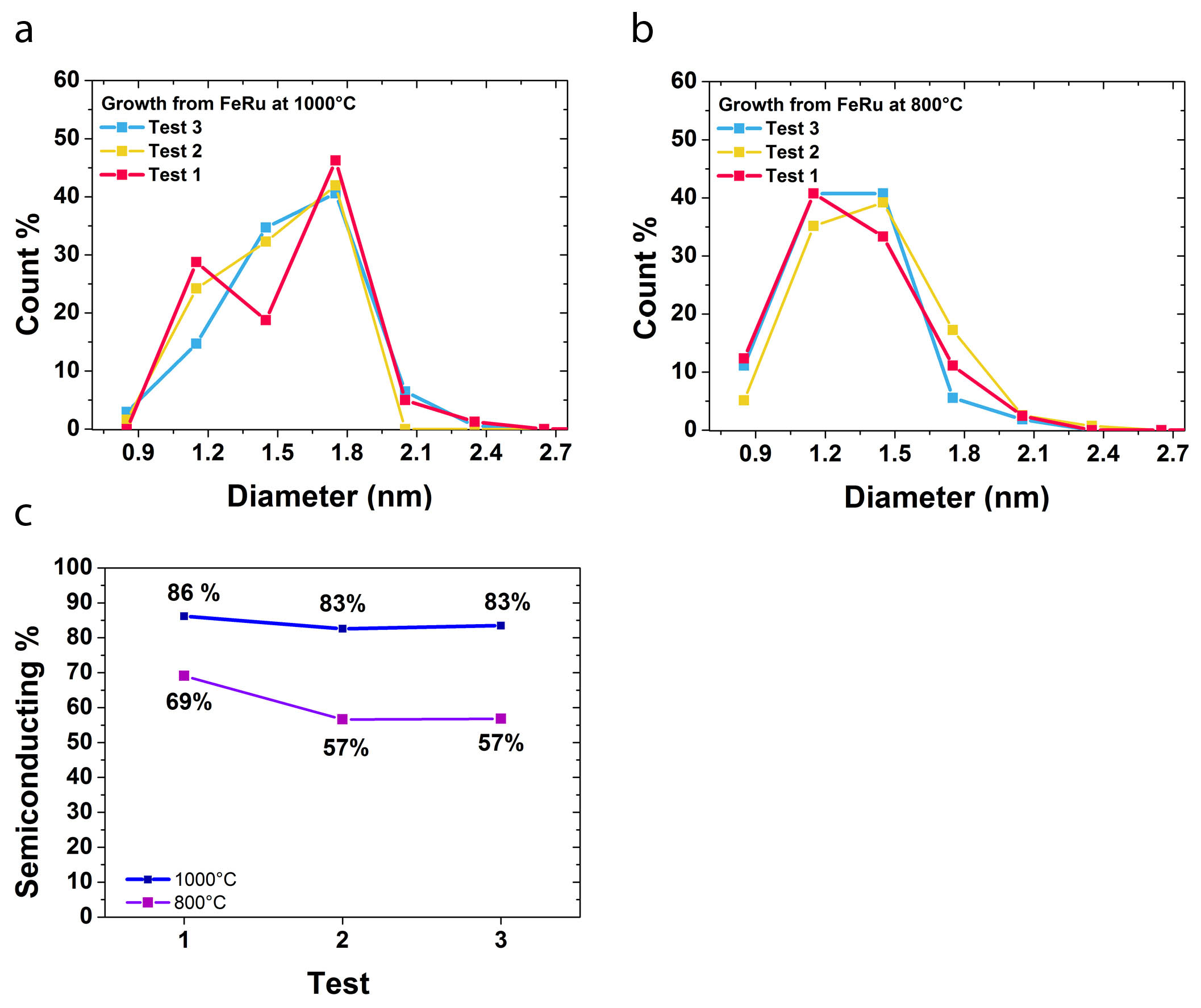}	
	\caption{Repartition of the SWCNT diameter obtained from FeRu catalyst for three different synthesis at a 1000\degree C b 800\degree C c the percentage of semiconducting nanotubes obtained for three different synthesis at 800 and 1000\degree C The diameter distribution was here obtained through statistical Raman spectroscopy analysis with three wavelengths (532 nm, 633 nm, 473 nm).}
\end{figure}

\newpage

\section{Dp/Dt}

\begin{figure}[ht]
\begin{center}

	\includegraphics[scale=0.67]{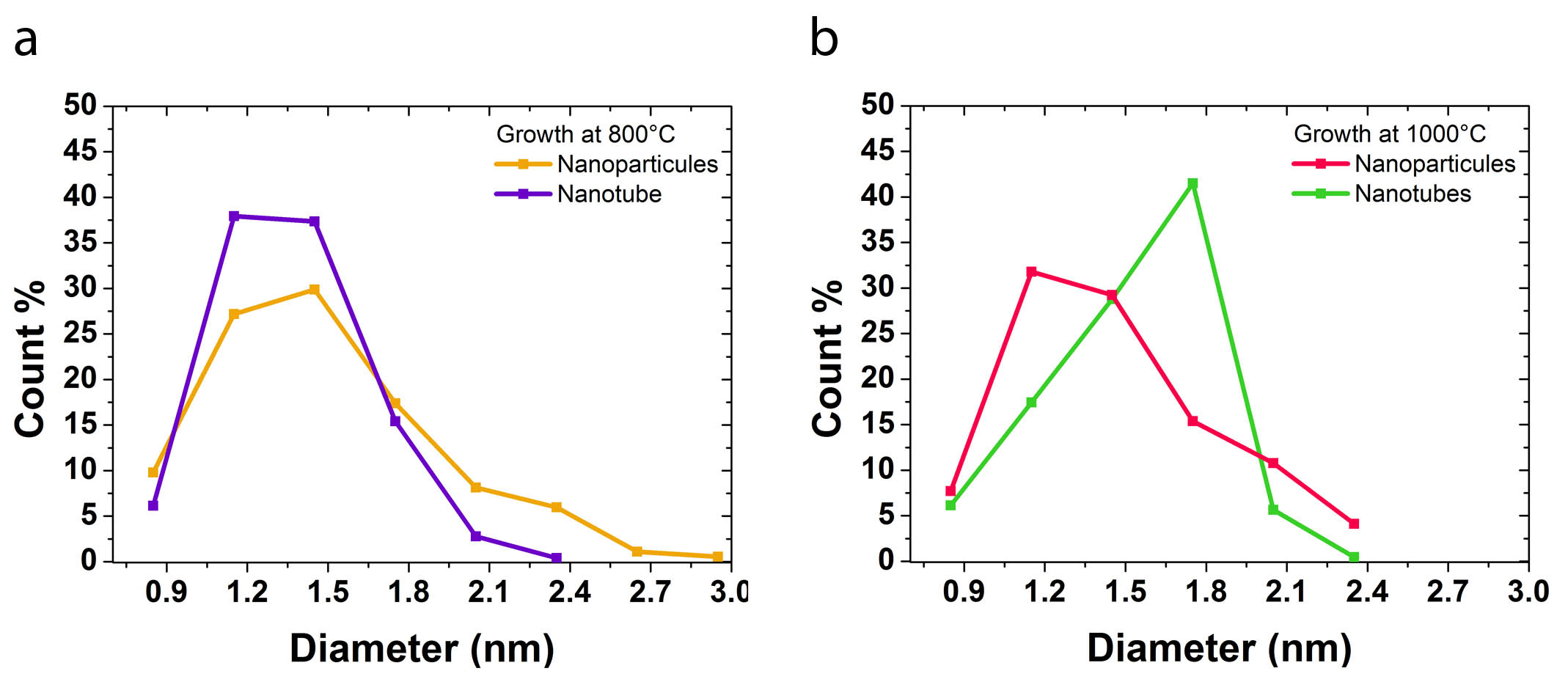}	
	\caption{Repartition of the diameter of FeRu catalyst and as grown SWCNT for synthesis at a 800\degree C and b 1000\degree C.}
		\end{center}
\end{figure}

\section{Bi-temperature synthesis}

\begin{figure}[ht!]
\begin{center}
	\includegraphics[scale=0.25]{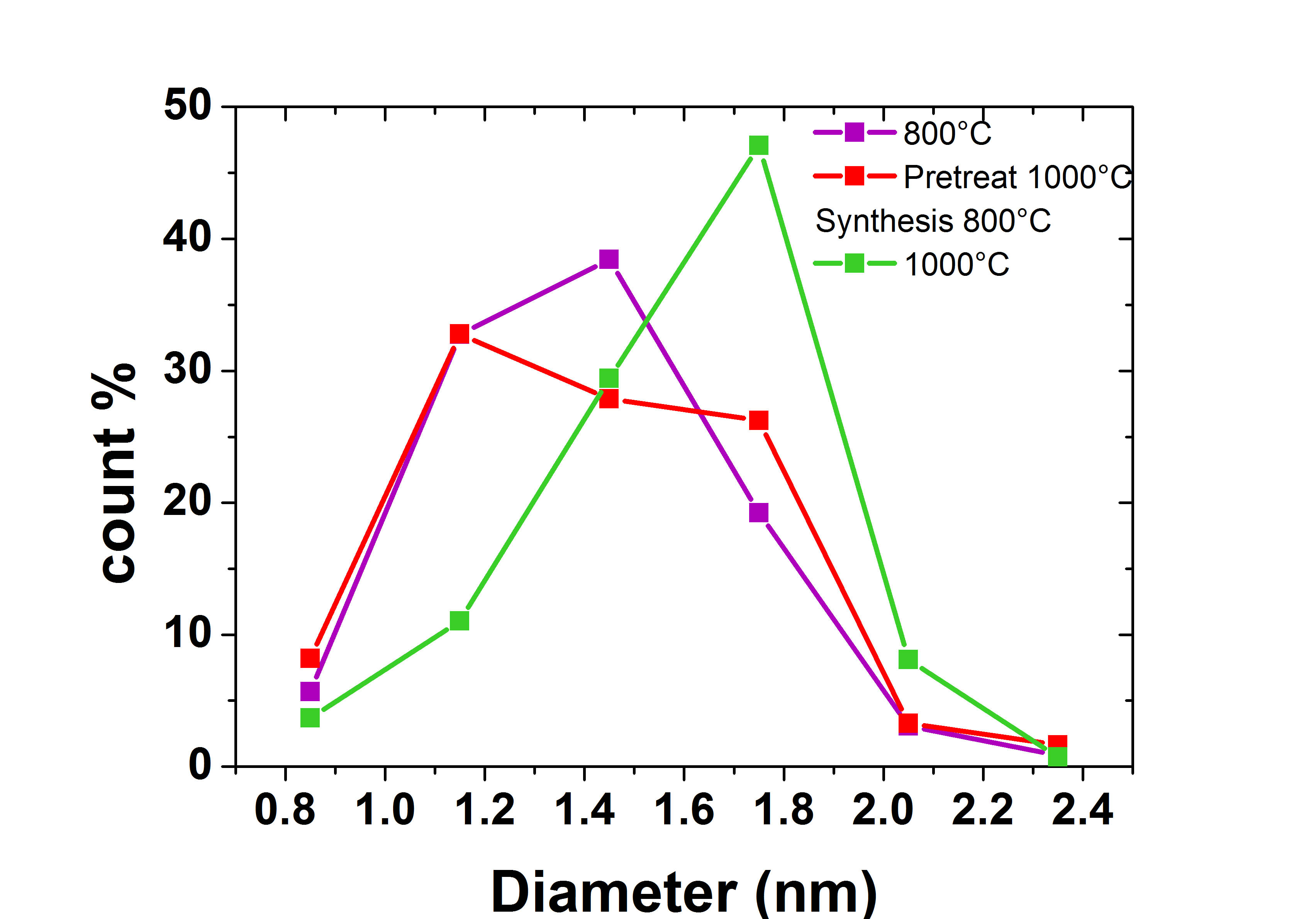}	
	\caption{size distribution histogram of nanotubes from FeRu catalyst for growth, at 800\degree C (pretreatment+growth) (purple line), pretreatment at 1000\degree C and growth
at 800\degree C (red line) and at 1000\degree C (pretreatment+growth) (green line). The diameter distribution was here obtained through statistical Raman spectroscopy analysis with three wavelengths (532 nm, 633 nm, 473 nm).}
	\end{center}
\end{figure}

\newpage

\section{SWCNT-FET}

\subsection{Geometry}

\begin{figure}[ht]
\begin{center}
	\includegraphics[scale=0.5]{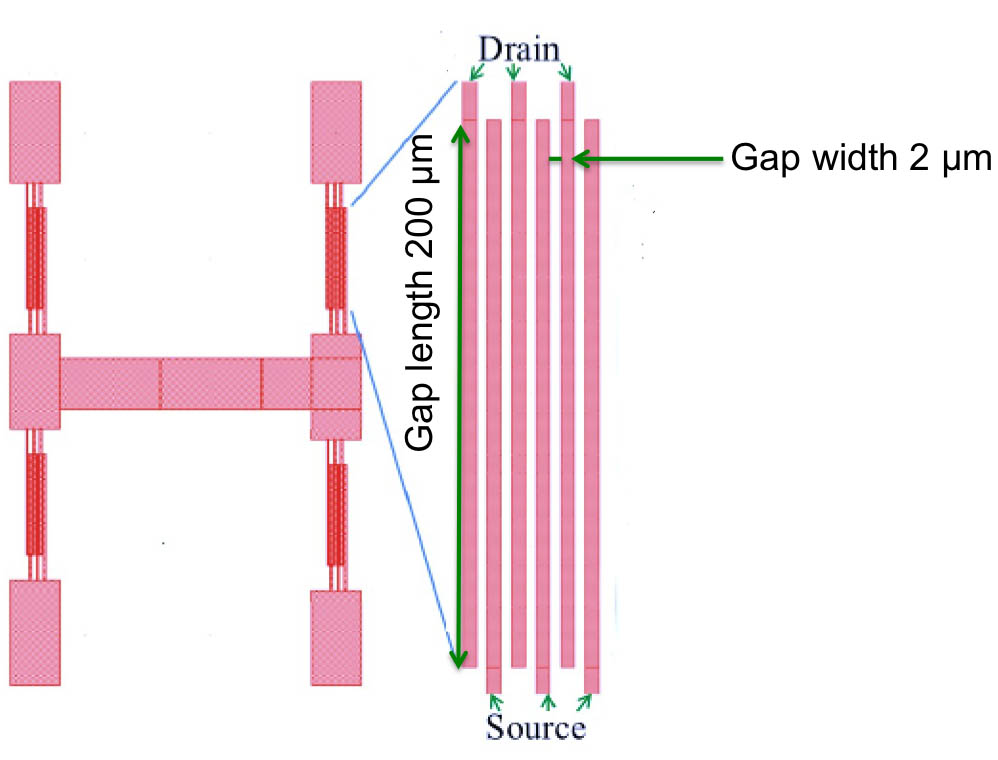}	
	\end{center}
	\caption{schematic view of the transistor geometry}
\end{figure}

\subsection{Characterization}

\begin{figure}[h!]
\includegraphics[width=\textwidth]{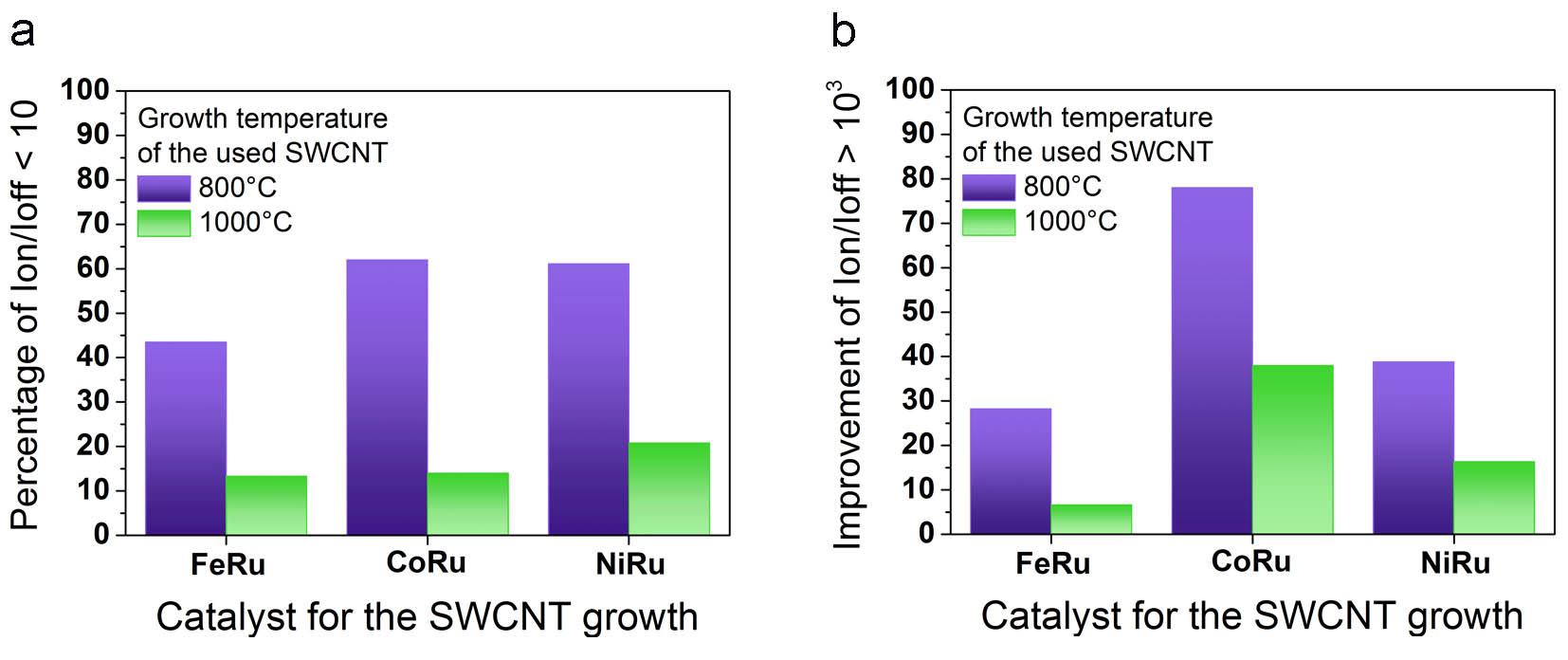}
\begin{center}
\begin{tabular}{c|ll}
catalyst & growth temperature & FWMH (excitation at 532nm)  \\ \hline  
 & 800 \degree C & 72.6 $\pm$ 2.4  \\
\multirow{-2}{*}{FeRu}   & 1000 \degree C & 79.9  $\pm$ 2.2 \\   \hline 
 & 800 \degree C &  59.8 $\pm$ 2.5   \\
\multirow{-2}{*}{CoRu}   & 1000
C & 63.9 $\pm$ 5.1 \\   \hline 
 & 800 \degree C & 71.4 $\pm$ 0.98 \\
\multirow{-2}{*}{NiRu}   & 1000 \degree C & 72.4 $\pm$ 2.66 \\   \hline 
  \end{tabular} 
\end{center}
\caption{\label{transistor-ineffective}   \textbf{a} Percentage of transistors presenting an Ion/Ioff ratio inferior at 10 for integrated SWCNT grown from each catalyst at 800 and 1000 \degree C (before electrical breakdown)  \textbf{b} Percentage of transistors presenting an improvement of their Ion/Ioff ratio superior at $10^3$ after electrical breakdown for integrated SWCNT grown from each catalyst at 800 and 1000 \degree C \textbf{Table} Table of the FWMH of the Raman D-band of samples grown from each catalyst at 800 or 1000 \degree C. }
\end{figure}

We also check that an increased presence of amorphous carbon present over the wafers' surface for synthesis at low temperature is not responsible for the higher percentage of ineffective transistor. In Raman spectroscopy, presence of amorphous carbon can be detected through the so-called D-band around 1350 $cm^{-1}$. Gao and co-worker, showed that for isolated SWCNTs, the full width half maximum (FWHM)  is between 20 $cm^{-1}$ up to 40 $cm^{-1}$  for an excitation at 632 nm \cite{Gao2008}. Picher and co-worker \cite{Picher2009}, for an excitation at 532 nm, reported a FWHM D-band between 50-60 $cm^{-1}$ for CVD SWCNTs comparable with the FWHM D-band of HiPco nanotubes in the same condition (FWMH 45 $cm^{-1}$). On the contrary they found a D-band FWHM in the range of 130-190 $cm^{-1}$ for amorphous carbon. In our synthesis, the full-width half-maximum of the D-band is found between 60-80 $cm^{-1}$ and no significative difference in the FWHM is observed for the different synthesis temperature for any of the studied catalysts (see table in Figure \ref{transistor-ineffective}).

\newpage

\begin{figure}[h!]
	\includegraphics[width=\linewidth]{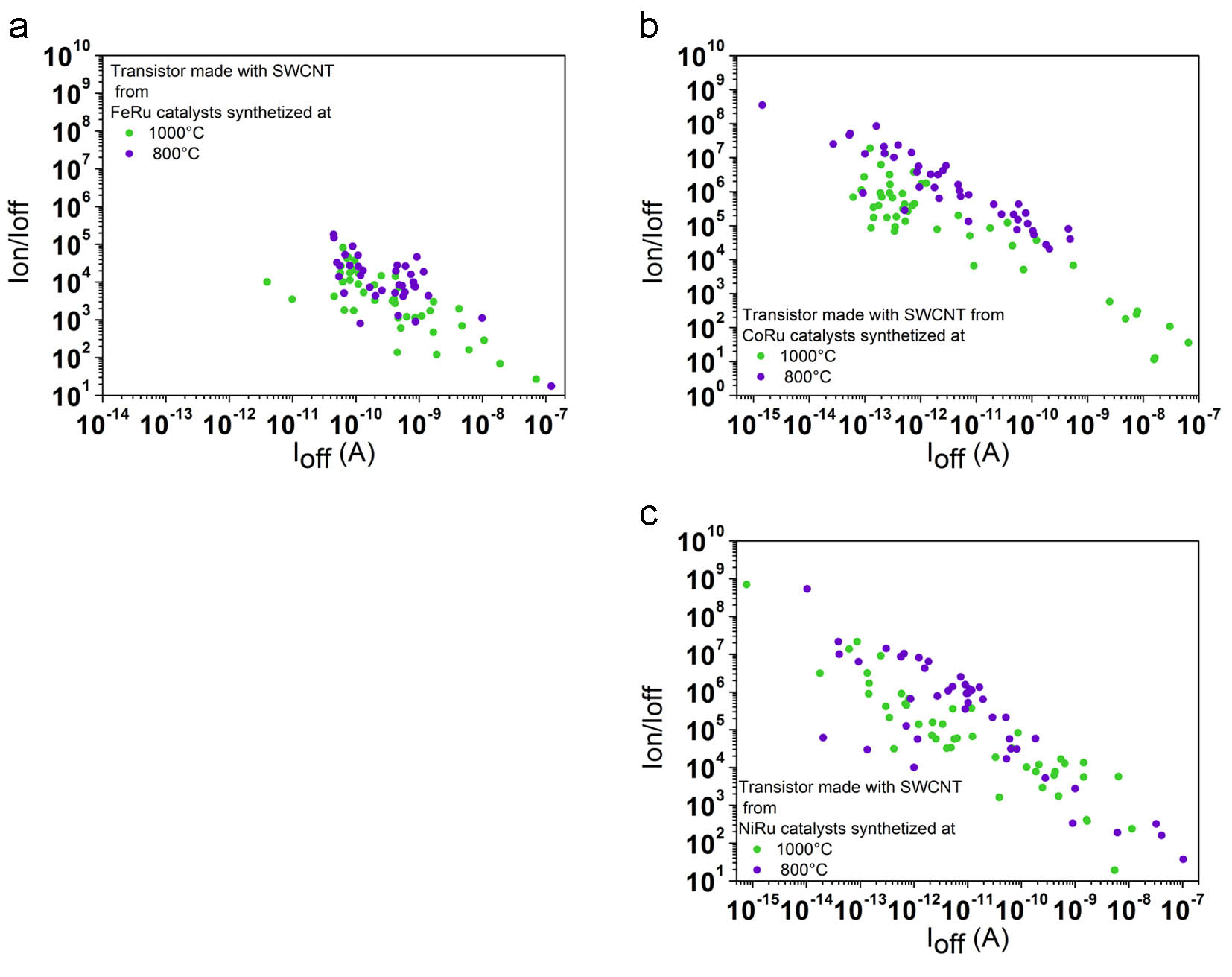}	
	\caption{Ion/Ioff ratio versus Ioff of FET-device after electrical breakdown fabricated from a FeRu catalyst b CoRu catalyst c NiRu catalyst}
\end{figure}

\newpage

\bibliographystyle{apsrev4-1}
\bibliography{biblio-swcnt-growth}